\documentclass[sigconf, screen, nonacm=true, authorversion=true]{acmart}
\usepackage{cleveref}
\usepackage{tcolorbox}
\usepackage{subfig}
\usepackage{graphicx}      
\usepackage{xspace}

\definecolor{aliceblue}{rgb}{0.94, 0.97, 1.0}
\tcbset{width=1\columnwidth,boxrule=1pt,colback=aliceblue,arc=1pt,left=2pt,right=2pt,boxsep=0.5pt}

\newcommand\mytodo[1]{\noindent{\color{orange} {\bf}{\bf }}}

\newcommand\TheName{Cascade\xspace}

\Crefname{figure}{Fig.}{Figs.}
\crefname{figure}{Fig.}{Figs.}
\crefformat{section}{#2§#1#3}
\crefformat{subsection}{#2\S#1#3}
\crefformat{subsubsection}{#2\S#1#3}
\AtBeginDocument{%
  }

\setcopyright{none} 
\copyrightyear{2025}
\acmYear{2025}
\acmDOI{XXXXXXX.XXXXXXX}

\acmConference[MICRO 2025]{The 58th IEEE/ACM International Symposium on Microarchitecture}{October 18--22, 2025}{Seoul, Korea}

\acmISBN{978-X-XXXX-XXXX-X/XX/XX}

\begin{document}





\newcommand{\fix}[1]{\color{red}{#1} \color{black}}
 \newcommand{\mqignore}[1]{}



\title{Utility-Driven Speculative Decoding for Mixture-of-Experts}


\author{Anish Saxena}
\email{asaxena317@gatech.edu}
\affiliation{%
  \institution{Georgia Tech, USA}
  \country{}
}

\author{Po-An Tsai}
\email{poant@nvidia.com}
\affiliation{%
  \institution{NVIDIA Research, USA}
  \country{}
}

\author{Hritvik Taneja}
\email{htaneja3@gatech.edu}
\affiliation{%
  \institution{Georgia Tech, USA}
  \country{}
}

\author{Aamer Jaleel}
\email{ajaleel@nvidia.com}
\affiliation{%
  \institution{NVIDIA Research, USA}
  \country{}
  }

\author{Moinuddin Qureshi}
\email{moin@gatech.edu}
\affiliation{%
  \institution{Georgia Tech, USA}
  \country{}
  }


\begin{abstract}
The GPU memory bandwidth is the primary bottleneck in low-latency Large Language Model (LLM) inference.
Speculative decoding leverages the underused GPU compute by employing a lightweight \textit{drafter} to propose $K$ tokens, which the LLM \textit{verifies} in parallel, increasing the token throughput.
Conventional \textit{dense} LLMs fetch \textit{all} model weights on each iteration, so there is no latency overhead due to speculation. 
The emerging Mixture of Experts (MoE) architecture activates only a subset of weights per token, significantly reducing data movement. 
However, we show that speculation is ineffective for MoEs: draft tokens collectively activate more weights, increasing data movement and verification time by 2–3x. When token throughput gains fail to offset verification overhead, speculation causes severe slowdowns--up to 1.5x--rendering it infeasible.  
Even when beneficial, the optimal length $K$ depends on both the task and the model, and can vary dynamically between requests and iterations.
Consequently, despite widespread adoption in dense LLMs, speculation remains unusable in leading MoEs.

We present \TheName, a utility-driven framework that selectively enables speculation to prevent slowdown and dynamically tunes $K$ to accelerate MoE serving. It uses a lightweight metric--\textit{speculation utility}--defined as the ratio of token gains to verification cost. We observe utility exhibits iteration-level locality, allowing periodic decisions through brief \textit{test} and prolonged \textit{set} phases. For each request, \TheName disables speculation if utility falls below one during testing. When the utility exceeds one, \TheName tests several $K$-values and selects the utility-maximizing $K$ for the set phase. We implement \TheName in vLLM and evaluate it on five popular MoEs with code, math, extraction, and mixed workloads representative of real-world serving. \TheName limits slowdown to 5\% (compared to 1.5x) and improves throughput by 7–14\% compared to static $K$-values, making speculative decoding practical for MoEs.

\end{abstract}

    \maketitle 

\section{Introduction}


\mytodo{state iso-accuracy somewhere.}
Large language model (LLM) inference and training dominate compute cycles, and reducing compute and data-movement overheads without sacrificing quality is a key challenge. Mixture-of-Experts (MoE)~\cite{shazeer2017moe}, a state-of-the-art LLM architecture, reduces data movement by sparsely activating a subset of parameters per token, enabling high-quality outputs at a fraction of the cost compared to dense models. Moreover, only the \textit{active} subset of model parameters are fetched for a token
Therefore, MoEs are suitable for low-latency serving where GPU memory bandwidth is the primary bottleneck. 
Leading LLMs, including Llama-4\cite{llama4}, GPT-4~\cite{gpt4_moe, gpt4report}, Gemini-1.5~\cite{gemini1.5}, and DeepSeek R1~\cite{guo2025deepseek}, leverage MoE architectures to scale effectively while maintaining state-of-the-art performance.


LLM inference for a request comprises a \textit{prefill} phase to process input tokens and \textit{decode} phase to generate output tokens \textit{autoregressively}.
The compute intensity of prefill phase is high as all prompt tokens of the request are processed in parallel.
In contrast, each decoding step requires fetching the billions of \textit{active} model parameters to process the next output token.
As the typical output length is hundreds of tokens, decoding time dominates the end-to-end latency of inference.
\mytodo{break this down, don't want to distract the limitation of single-batch serving from the focus.}
Moreover, in single-batch serving, which is the focus of this work, the compute units are underutilized, and the execution time of each decoding step is governed by data movement to fetch the model parameters. As a result, LLM inference is highly memory bandwidth bound owing to high data-movement costs.

\begin{figure*}
    \centering
     \includegraphics[width=0.95\textwidth]{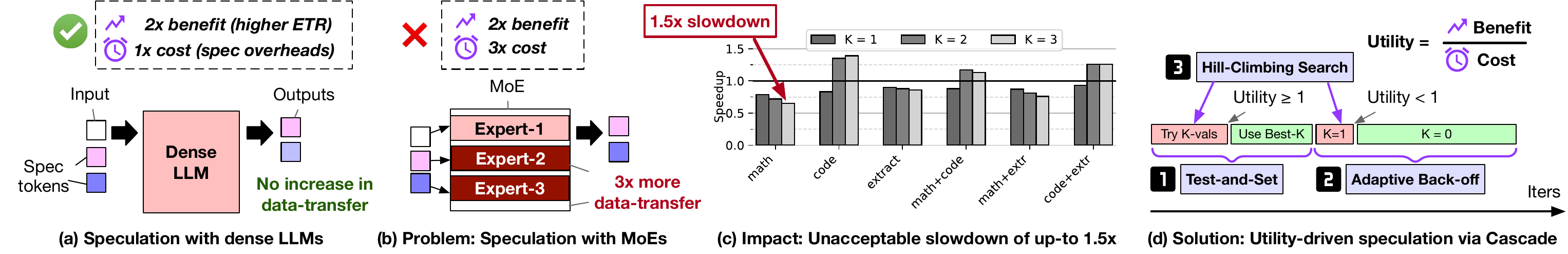}    
      \vspace{-3mm}
     \caption{(a) Speculation in dense LLMs incurs no added memory traffic (b) Speculation in MoEs increases data-movement during verification by 2x-3x (c) Performance of static-$K$ N-gram speculation in Mixtral (up to 1.5x slowdown) (d) \TheName: (1) continuous monitoring via test-and-set, (2) adaptive backs off to minimize test cost, and (3) hill-climbing to find utility-maximizing $K$.} 
    \label{fig:intro}
     \vspace{-3mm}
\end{figure*}

Speculative decoding~\cite{speculative-decoding, chen2023accelerating} accelerates LLM serving by leveraging underutilized GPU compute to emit multiple tokens per decoding step. A lightweight \textit{drafter} generates $K$ draft tokens before each iteration of the \textit{target} model, which verifies the tokens in parallel. 
With accurate drafts, the effect token rate (ETR) increases. Drafters run 50x–100x faster than the target model and incur minimal overhead.
The memory bandwidth bottlenecks verification latency, which remains unchanged in \textit{dense} LLMs that load all model weights every iteration. As Figure~\ref{fig:intro}-(a) shows, speculation~\cite{speculative-decoding, cai2024medusa, li2024eagle} increases ETR \textit{without increasing memory pressure}, improving performance. In our evaluations with LLaMA-3-8B~\cite {grattafiori2024llama}, speculation~\cite{saxena2023prompt} provides consistent speedup of 1.4-1.8x.

In this work, we show that speculation is \textit{not} a feasible solution for MoEs. 
\mytodo{typically, spec is effective for dense models because...}
Speculation is effective because the time to verify draft tokens remains unchanged compared to decoding a single token, making even marginal improvements in ETR beneficial.
However, in MoEs, each token independently selects a subset of experts, leading to higher data movement due to speculation. 
For example, in Mixtral~\cite{jiang2024mixtral}, a single token activates only 2 out of 8 experts for each layer.
With three draft tokens, up to 6 experts may be activated, leading to a 3x increase in data movement to fetch the experts. 
Speculation will not be beneficial unless ETR improves by more than 3x.
As Figure~\ref{fig:intro}-(b) illustrates, we find that the verification overhead is often greater than the improvement in ETR. 

Consider Figure~\ref{fig:intro}-(c), where we show the performance of Mixtral~\cite{jiang2024mixtral} MoE across speculation lengths $K \in {1, 2, 3}$ on tasks such as code, math, natural language extraction, and mixed workloads that reflect the heterogeneity of real-world serving. We find that \textit{every} workload experiences slowdown for at least one $K$, and three (math, extract, math+extract) suffer slowdown across \textit{all} $K$-values. Notably, a $K$-value that performs well for one task can induce severe degradation on others--up to 1.5x in our evaluation. These results underscore a key limitation: profiling an optimal static-$K$ per task is ineffective under realistic, mixed-request serving scenarios.


\noindent\textit{The Problem:}
Speculation--a latency-reducing technique--is typically \textit{always-on} due to low overheads, but doing so incurs unacceptable slowdown with MoEs, rendering it unusable for leading LLMs.

Ideally, when speculation is detrimental, we want to predict the \textit{no-speculation} state to disable it dynamically. Prior schemes tune $K$ for dense models~\cite{brown2024ddd,zhang2024svip,mamou2024disco,liu2024pearl} by leveraging the drafter's probability distribution. However, these schemes maximize ETR without being cost-aware, requiring the generation and verification of at least one draft token.
Thus, such schemes cannot turn off speculation altogether. 
As Figure~\ref{fig:intro} illustrates, even a conservative $K=1$ can incur a performance loss exceeding 25\% (for the math task), limiting their appeal. 
Moreover, the drafter's ETR is variable due to inherent token prediction difficulty.
The verification costs also fluctuate, partly from changing temporal affinity between successive tokens and experts~\cite{hwang2024pregated, huang2023towards}. 
Consequently, the optimal $K$ depends on the task, model architecture, and varies even between requests and iterations within the same task~\cite{mamou2024disco}.

\noindent\textit{Our Goal:} We want to develop a lightweight scheme that adaptively disables speculation if necessary and when it is beneficial, tunes $K$ dynamically to maximize speculation efficacy for MoEs.

In this paper, we present \TheName, a utility-driven speculative decoding framework, which selectively applies speculation to accelerate MoE serving. We introduce a simple heuristic, the \textit{utility of speculation}, defined as the ratio of benefits of speculation--improvement in ETR--to its costs--the verification overhead. For example, if speculation improves the ETR by 1.5x but increases verification time by 2x, the utility is 0.75. 
We show that utility robustly predicts speculation performance and provides a simple knob to turn-off speculation (when utility is less than one). 
Moreover, we prove that maximizing utility directly minimizes the average time per output token (TPOT), a key performance metric.
For each request, \TheName tracks recent speculation costs and benefits, computes runtime utility, and selects utility-maximizing $K$ (including $K=0$) to improve future utility of speculation.
As Figure~\ref{fig:intro}-(d) shows, our utility analysis provides three key insights guiding our design.

\mytodo{flow is not coming across. why the next 3 paragraphs? improve flow.}
\mytodo{more numbers. what's the impact of each thing? less information. Talk about spec manager and utility analyzer and how they help. More Midas}
\mytodo{Instead of we must do this, say Midas ensures/ does this.}
\textit{Test-and-Set to Maximize Utility.} We observe that utility, though dynamic, exhibits temporal locality across short iteration intervals. Therefore, we employ a \textit{test-and-set} policy to monitor utility and tune $K$ periodically. In the test phase, we evaluate up to four distinct $K$-values for four iterations each and measure their utility. The test phase is sized to explore multiple $K$-values while efficiently obtaining reliable utility estimates. Next, we select the $K$ value that maximizes utility for the subsequent set phase, which spans 16 iterations. When the test-phase utility is below one, even at the minimal $K=1$, we disable speculation during the set phase to prevent performance degradation. Our test-and-set interval is short (32 iterations), allowing rapid adaptation to changes in utility.

\textit{Adaptive Back-off to Minimize Testing Cost.}
We observe that requests experience phases of high and low utility, with some initially exhibiting low utility that improves later in the generation process. Thus, periodic testing is essential to avoid making decisions on speculation for the request's lifetime. However, even when speculation is disabled during the set phase, testing costs can be substantial. For example, if the current utility is 0.5 (or 2x slowdown), using 4 out of 20 iterations for testing would incur a 20\% slowdown. To mitigate this overhead, we amortize testing costs by doubling the set-phase duration whenever we switch to the $K=0$ state. During subsequent test phases, we continue exploring various $K$ values to capture opportunities to speculate effectively. Adaptively backing off reduces \TheName's worst-case slowdown from 14\% to just 5\%.

\textit{Hill-Climbing to Find Utility-Maximizing $K$.} 
We must carefully select the $K$-values during the testing phase, since suboptimal choices can degrade performance or even induce slowdown.
We observe that the \textit{direction} of change in utility provides clues to choosing the utility-maximizing $K$ (or best-$K$).
For instance, if utility decreases by increasing $K$, the costs of verifying more tokens likely dominate the benefits of improved ETR, and we should backtrack to a lower $K$.
Therefore, we employ a hill-climbing scheme that uses recent $K$ and utility trends to guide $K$ selection during testing, and adopts the best-$K$ found for the subsequent set phase.
While profiling static-$K$ for each task is impractical owing to mixed request streams, \TheName matches or exceeds the best static-$K$ across all evaluated tasks, achieving up to 14\% average performance gain for Mixtral.

We implement \TheName on the vLLM serving framework~\cite{pagedattention_vllm} and evaluate on 5 popular MoEs and 7 tasks, including mixed requests. \TheName provides performance improvement of 20\% on average (compared to a non-speculative baseline) when speculation is beneficial, and incurs minimal slowdown of at-most 5\% if speculation is ineffective, thereby making speculation practical for MoEs. 

In summary, this paper makes the following contributions:

\begin{itemize}
    \item We show that unlike dense models, speculation is not practical for MoEs due to higher data-movement in verification.
    \item We find that speculation utility is a robust predictor of speculation performance, and an effective guiding metric.
    \item We present, \TheName, a utility-driven speculation framework to dynamically disable speculation if it is detrimental.
    \item We limit worst-case slowdown of speculation to just 5\%, compared to prohibitive 54\% loss in static-K schemes.
    \item We propose a hill-climbing search to find utility-maximizing $K$ adaptively, outperforming static-$K$ schemes by 7-14\%.
\end{itemize}



\mqignore{
Disabling speculation when it is detrimental improves performance marginally, yielding an average speedup of XY\% across tasks. In addition to  effective approach would be to reduce the verification overhead of speculative tokens, allowing speculation to be applied more frequently. We observe that in 40-70\% of iterations, speculation is incorrect and does not yield any additional tokens. To this end, our key insight is to identify misspeculated tokens early, within intermediate layers—for instance, by layer 8 out of 32. Once detected, these tokens are pruned from further computation, preventing them from propagating through the remaining layers (layers 9 through 32). This ensures that only the always-accepted token is passed through all layers, reducing the verification overhead by 75\%. This technique, which we term dynamic early-exit, preserves output quality. Moreover, it reduces the average verification cost to X-Yx, leading to an average speedup of XY\% across tasks over a non-speculative baseline.
}

\mqignore{
Reducing verification cost when speculation is beneficial requires fetching fewer experts than those selected by the gating network, without compromising output quality. To achieve this, we introduce an adaptive expert restriction mechanism that dynamically limits the set of experts loaded during verification. We take a conservative approach, prioritizing correctness by aggressively rejecting uncertain draft tokens rather than accepting incorrect ones. Specifically, we always retain experts required by the always-taken token to ensure baseline correctness. Additionally, since expert activations are weighted before aggregation, we selectively retain a subset of experts that capture the majority of activation information while discarding those with minimal contribution. Our evaluations show that this method accepts and rejects the same tokens as the baseline, with negligible quality degradation (1-2\%). Furthermore, by capping the number of experts per iteration to at most X, our approach reduces verification overhead to Xx, yielding an XY\% average speedup across tasks compared to a non-speculative baseline.
}

\newpage

\section{Background and Motivation}

\subsection{LLM Architectures: Dense versus MoE}

\begin{figure}[h]
    \centering    
           \includegraphics[width=.95\columnwidth]{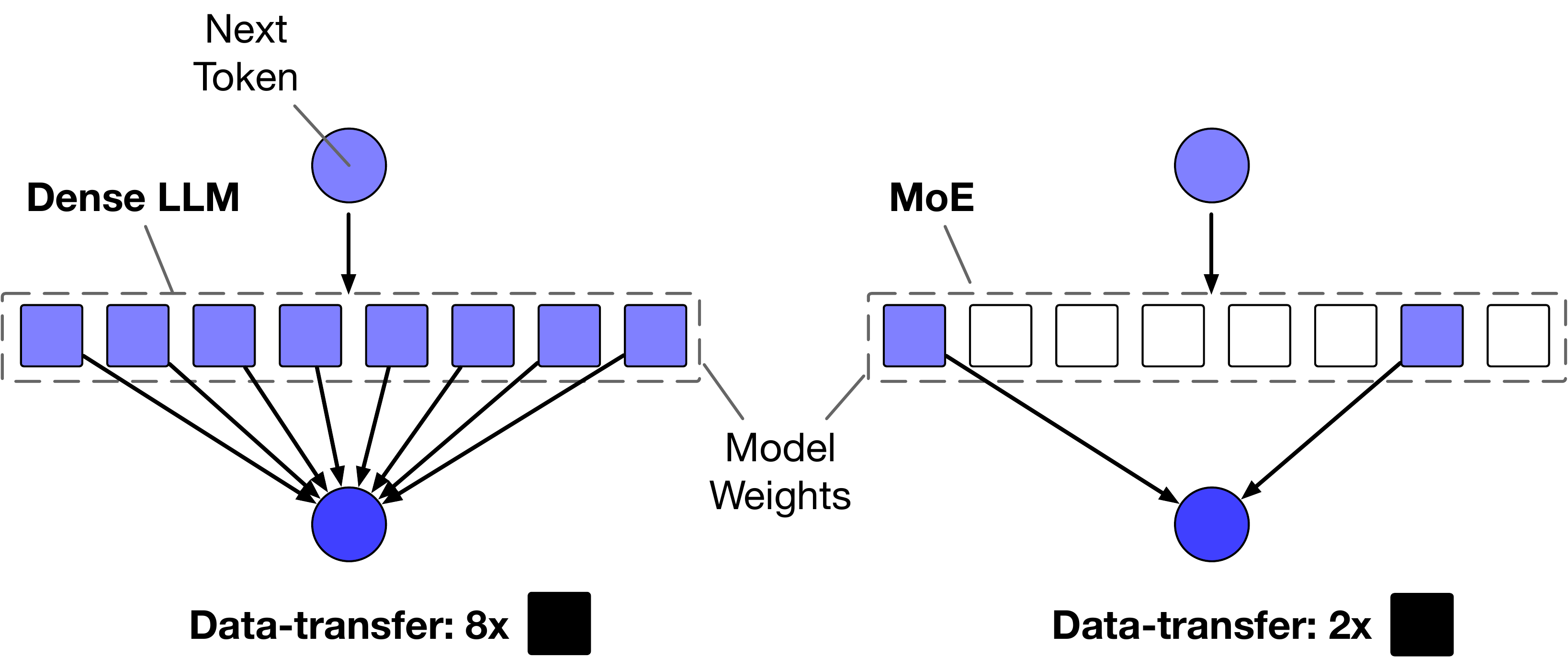}
    \caption{Unlike dense models, MoEs activate only a subset of the model, significantly reducing the data-movement.}
    \label{fig:moe_overview}
\end{figure}

Large Language Models (LLMs)~\cite{gpt3paper} excel in a wide range of tasks, including math, code, and even multi-model queries~\cite{gpt4report}.
An LLM consists of multiple transformer blocks~\cite{vaswani2017attention}, each composed of self-attention and feedforward layers that dominate execution time. Inference proceeds in two phases: prefill, which processes input tokens in parallel to generate the first output token, and decode, which generates output tokens autoregressively. Decode often requires hundreds of steps for long outputs and dominates latency. Each decode step fetches billions of active parameters from GPU memory, making it memory bandwidth-bound. In conventional dense LLMs, even a single token activates \textit{all} model weights (Figure~\ref{fig:moe_overview}-left). As model size grows, this data movement imposes severe bandwidth and capacity bottlenecks on low-latency serving.

\smallskip
\noindent\textbf{Mixture-of-Experts.} The sparse Mixture-of-Experts (MoE) architecture improves compute efficiency during both training and serving by activating only a small portion of the model per token. Each transformer block includes multiple feedforward \textit{experts}, and a lightweight \textit{router}—a small gating network—selects a few experts to process each token. Recent publicly available MoE models match or exceed dense model quality, and commercial LLMs such as GPT-4, LLaMA-4, DeepSeek R1, and Gemini 1.5 use MoE architectures~\cite{jiang2024mixtral, switchtransformer, dbrx, nllbteam2022language, qwen_moe, gpt4_moe, snowflake_arctic, llama4}. MoEs reduce data movement significantly by fetching only the selected experts. For example, in Figure~\ref{fig:moe_overview}-right, the router activates only 2 out of 8 experts.

\mqignore{
\subsection{LLM Inference}


LLM inference is predominantly memory-bound and consists of two phases: prefill and decode. During the prefill phase, input tokens for each sequence in the batch are processed in parallel, passing through the model to generate the first output token. This phase is compute-intensive, as it processes hundreds of input tokens simultaneously to construct the prompt context (KV-states) for each sequence. In contrast, the decode phase generates one token at a time per sequence, progressing sequentially due to the autoregressive nature of LLMs. As a result, generating hundreds of output tokens per sequence requires hundreds of decode steps.


\smallskip
\noindent\textbf{LLM Decoding.} At each decode step, active model parameters are fetched from memory to compute a single token per sequence, reducing the FFN layers' GEMMs to memory-bound matrix--vector operations. Consequently, the operational intensity (OI) of decoding—measured as the ratio of compute FLOPS to data movement from memory-—is 500x lower than the compute-intensive "knee" of modern GPUs. While increasing batch size or "chunking" the prefill tokens of some sequences with decode steps of other sequences [CITE] improves OI, it is not always feasible. In edge deployments, models typically serve a single sequence at a time, and in latency-critical serving, waiting for requests to form larger batches violates latency SLOs. As a result, single-batch serving remains highly memory-bound. For instance, X\% of end-to-end latency is spent in the decode phase for tasks like math, code, and information extraction [CITE] in the LLAMA-3-8B dense model at BS = 1. 
}

\subsection{Speculative Decoding}
\smallskip
\noindent\textbf{Speculative Decoding.} 
\mytodo{avoid too many double hyphens.}
Speculative decoding\cite{speculative-decoding} improves LLM inference throughput by emitting multiple tokens per decoding step, significantly reducing time per output token (TPOT), a key performance metric. A lightweight \textit{drafter}--10x–100x faster than the target model--generates $K$ draft tokens. The main \textit{target} model then \textit{verifies} the next-token prediction from the previous decoding step and the draft tokens in a single step. Finally, a rejection sampler accepts or rejects the draft tokens. Although verification requires additional compute, its latency remains unchanged due to the memory-bound nature of LLM inference. Speculation yields up to $K{+}1$ tokens per iteration, since the system always emits at least one token. Drafters span a wide spectrum, from n-gram matching methods\cite{saxena2023prompt} to learned models like EAGLE~\cite{li2024eagle}, which autoregressively predict future hidden states.

\begin{figure}[h]
    \centering    
           \includegraphics[width=.95\columnwidth]{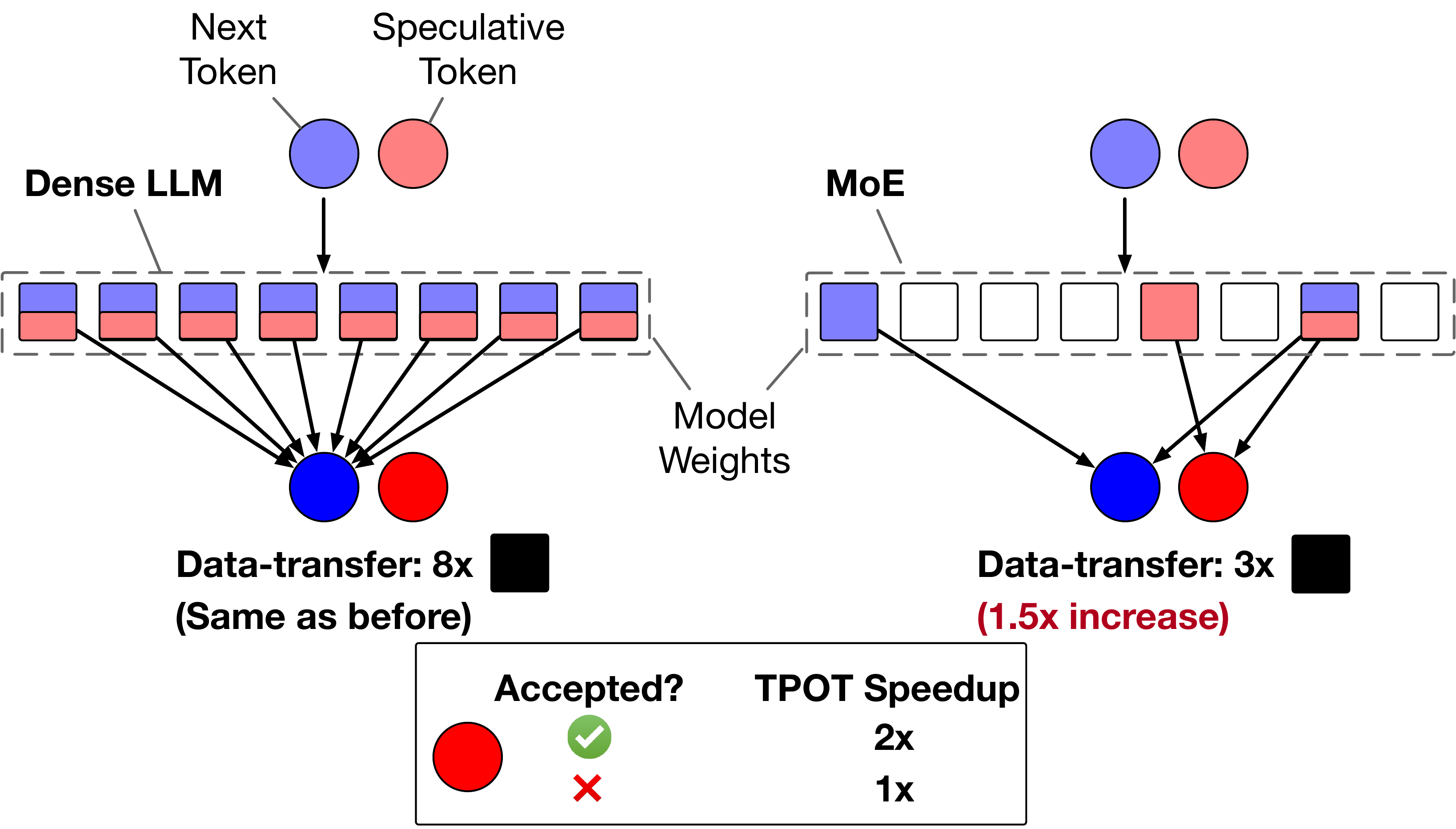}
    \caption{Speculation improves TPOT without increasing memory pressure in dense models. In MoEs, speculation incurs additional data-movement as more experts are activated.}
    \label{fig:spec_decode_overview}
\end{figure}

Speculative decoding exploits underutilized compute during decoding \textit{without increasing memory bandwidth pressure}. This strategy works well for dense models, which fetch all model parameters at each iteration regardless of the number of tokens processed. In MoEs, each token only activates a subset of experts. However, during verification, speculative tokens may trigger different expert subsets, collectively activating a larger fraction of the model. As a result, speculation increases data movement in MoEs, as shown in Figure~\ref{fig:spec_decode_overview}. Since execution latency is governed by the time to fetch active parameters from GPU memory, the increased activation increases memory bandwidth pressure. If the resulting gains in output tokens fail to offset the added cost, speculation degrades performance--a behavior we analyze next.

\mqignore{
\subsection{Speculative Decoding Techniques}


\smallskip
\noindent\textbf{Speculation by Matching N-Grams.}
For input-grounded tasks such as summarization, code generation, and conversations, a lightweight drafting method is n-gram speculation, which uses simple string matching to generate candidate tokens. This eliminates the need for a separate draft model, making it model-agnostic and compatible with greedy or sampling-based decoding. At each drafting step, the most-recent N tokens are matched against input context n-grams; upon a successful match, the next K tokens in-context become proposal candidates. N-gram offers a low-overhead, architecture-independent approach to accelerate decoding.

\smallskip
\noindent\textbf{Speculation using Draft Models.}
Employing a small draft model to propose tokens requires shared tokenizers and vocabularies with the target model, for compatibility with rejection sampling. Finding a suitable off-the-shelf draft model is challenging, as it must be significantly smaller (10x-100x) yet retain the target's large vocabulary. Recent approaches address this by augmenting the target model or explicitly training inexpensive drafters. Medusa, for example, integrates extra decoding heads for parallel token prediction via tree-based attention. Similarly, EAGLE uses a feature-level autoregressive draft model predicting future hidden states for token generation. These methods outperform n-gram in speculation accuracy, but require dedicated drafter training and complicate LLM architecture and distributed serving [CITE].
}


\mqignore{
\subsection{Cost-Benefit Tradeoff of Speculation}

\mytodo{ Only need one small paragraph about ETR and benefit. Don't need to expand on accept rate.}

\smallskip
\noindent\textbf{Benefit of Speculation.}
A model-level metric for speculation is the \textit{acceptance rate}, which is the per-token probability of a proposal being accepted by the target model. 
A high acceptance rate indicates that the draft model effectively mimics the target model's output distribution. 
However, acceptance rate does not capture the efficacy of speculation, as it does not consider causal acceptance. 
For instance, for $K=3$ speculative tokens with rejection-sampling output of $[Acc,Rej,Acc]$ for proposed tokens, the acceptance rate would be 0.67, even though only one token is accepted (causal-acceptance rate of 0.33).
Moreover, acceptance rate does not account for the \textit{always-taken} token, which is output even if all proposals are rejected.
From a systems standpoint, a more relevant metric is the \textit{effective token rate}, which we define below.

\mytodo{ETR not being best metric is not a debate we really need to have.}

\begin{definition}
\label{def:etr}
\textbf{Effective Token Rate (ETR):} The average number of tokens output by the target model per iteration.
\end{definition}

We compute ETR by simply summing up the number of tokens output by the target model across all iterations, and dividing it by the total number of iterations.
For typical speculation techniques, the effective tokens that are output every iteration (ETR), which represents the \textit{benefit} of speculation, is greater than one.
\cref{fig:llama_spec_iter_time}-top shows the improvement in ETR and TPOT achieved by speculative decoding techniques for the dense LLAMA-3-8B model on math, code, and extraction tasks, normalized to non-speculative baselines.
A higher ETR indicates that the target model outputs more tokens per iteration, leading to higher throughput, and ETR speedup reliably predicts TPOT improvement.
Overall, speculation improve performance by 1.2x-1.8x compared to the non-speculative case, without causing TPOT degradation in any task.
}



\subsection{The Effectiveness of Speculation}

\begin{figure*}[htb!]
\centering
\subfloat{%
    \includegraphics[width=1.95\columnwidth]{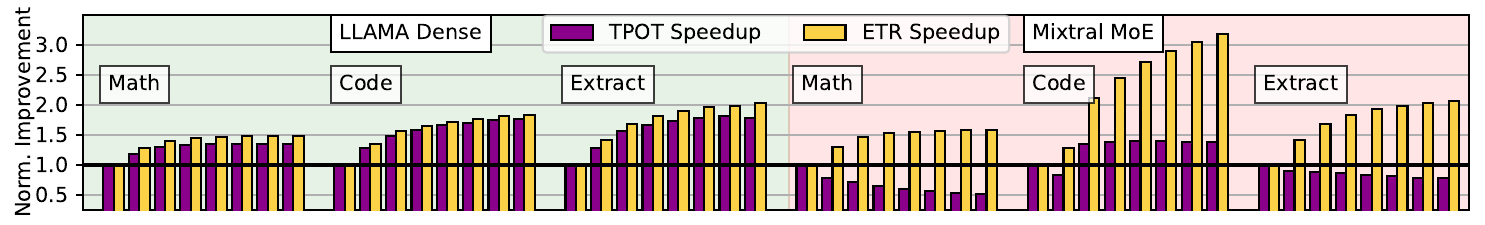}
}
\vspace{-2mm}
\hfill
\subfloat{%
    \includegraphics[width=1.95\columnwidth]{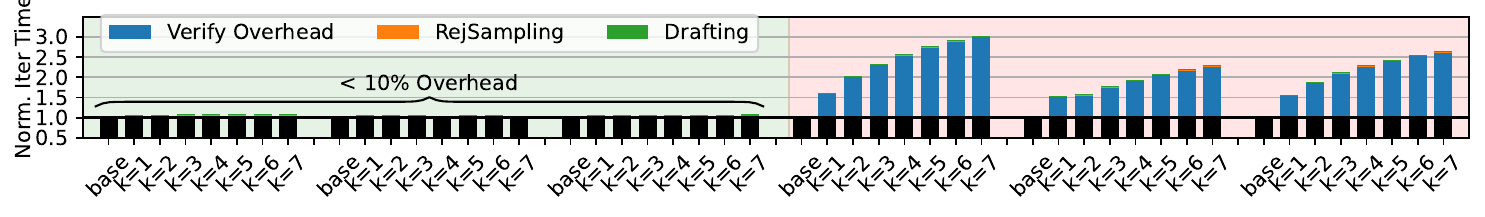}
}
\vspace{-2mm}
\caption{Top: TPOT and ETR speedup with n-gram speculation on dense LLAMA-3 (left, green) and Mixtral MoE (right, red) models. Bottom: Speculation iteration time breakdown. In dense model, speculation overhead is low at 5-10\%. However, speculation imposes significant overheads of 1.5x-3x for MoE, and several scenarios exhibit drastic TPOT slowdown of upto 2x.
}
\label{fig:spec_overheads}
\end{figure*}

\mytodo{highlight ETR definition.}
The primary benefit of speculation is a higher effective token rate (ETR)—the average number of tokens the target model emits per iteration, which improves TPOT (time per output token). Figure~\ref{fig:spec_overheads}-top shows ETR and TPOT improvements for n-gram speculation on the dense LLaMA-3-8B model (green), with speculation lengths $K \in [1, 7]$ (evaluation details in Section~\ref{sec:exp_setup}). As $K$ increases, ETR rises by 1.5x-2x, yielding a TPOT speedup of 1.4x–1.8x at $K=7$.


The \textit{cost} of speculation is three-fold: 

\begin{enumerate}
    \item Drafting-overhead: the time required to generate speculative tokens by the draft model.
    \item Verification-overhead: the additional time to verify the speculative tokens by the target model.
    \item Rejection sampling-overhead: the time required to accept or reject the speculative tokens.
\end{enumerate}

A central premise of speculative decoding is that its overheads remain low. \Cref{fig:spec_overheads}-bottom breaks down iteration time into its major components. Across all configurations, speculation adds only 5\%–10\% overhead, primarily from rejection sampling and drafting. The verification overhead is negligible (maximum of 3\%) since dense models fetch all weights regardless of token count, keeping execution time unchanged. Because speculation incurs minimal cost, even modest ETR gains yield performance benefits. Consequently, systems typically keep speculation \textit{always-on}.






\smallskip
\noindent\textbf{Speculation with MoEs.}
\Cref{fig:spec_overheads}-top (red) shows the TPOT and ETR improvements from n-gram speculation on the Mixtral 8x7B MoE~\cite{jiang2024mixtral}. Like dense models, Mixtral achieves substantial ETR gains of 1.6x–3.2x at $K{=}7$ across tasks. However, TPOT speedup often fails to track ETR improvement. For example, despite a 3.2x increase in ETR for code, the TPOT speedup is only 1.4x. This mismatch makes speculative performance in MoEs unpredictable. More concerning, two tasks--math and extraction--suffer slowdowns for \textit{all} speculation lengths. Unlike dense models, where speculation consistently improves latency (\cref{fig:spec_overheads}-green), MoEs can experience severe performance degradation under speculation.



\mqignore{
\smallskip
\begin{tcolorbox}[boxrule=1pt,left=5pt,right=5pt,top=1.5pt,bottom=1.5pt]
    \textbf{Takeaway:} \textit{Unlike dense models, speculation can induce significant slowdown in MoEs, and ETR improvement fails to predict TPOT speedup reliabily.} 
\end{tcolorbox}
}

\Cref{fig:spec_overheads}-bottom shows the iteration time breakdown for Mixtral under speculation. Rejection sampling and drafting remain inexpensive, incurring only 1–2\% overhead. However, verification incurs a much higher cost in MoEs. As speculation length increases, verification overhead increases due to greater data movement. Unlike dense models, this overhead is not constant and varies by task, reaching 3x for math, 2.3x for code, and 2.6x for extraction at $K{=}7$.


\smallskip
\begin{tcolorbox}[boxrule=1pt,left=5pt,right=5pt,top=1.5pt,bottom=1.5pt]
    \textbf{Takeaway:} \textit{MoEs break the key assumption in speculative decoding that the verification overhead--the cost of verifying speculative tokens compared to decoding a single token--is negligible.} 
\end{tcolorbox}

\subsection{The Increase in Activated Experts}

During verification, attention accounts for only 8\% of baseline latency, and its cost remains stable across speculation lengths. In contrast, expert-layer latency grows with $K$-value as the router activates more experts. For example, at $K{=}7$, Mixtral processes eight tokens per step and selects two experts per token. Assuming uniform random selection of experts by the router, this leads to over seven unique expert activations, on average, a 3.5x increase in data movement, based on a bucket-and-balls analysis. In practice, the increase is lower due to \textit{expert affinity}~\cite{hwang2024pregated, huang2023towards}; consecutive tokens often reuse experts, improving reuse beyond random chance. For instance, the math task in \cref{fig:spec_overheads} shows only a 3x overhead at $K{=}7$. Although verification cost varies across tasks and depends on affinity, reducing $K$ consistently lowers overhead.

\begin{figure*}
    \centering    
           \includegraphics[width=1.95\columnwidth]{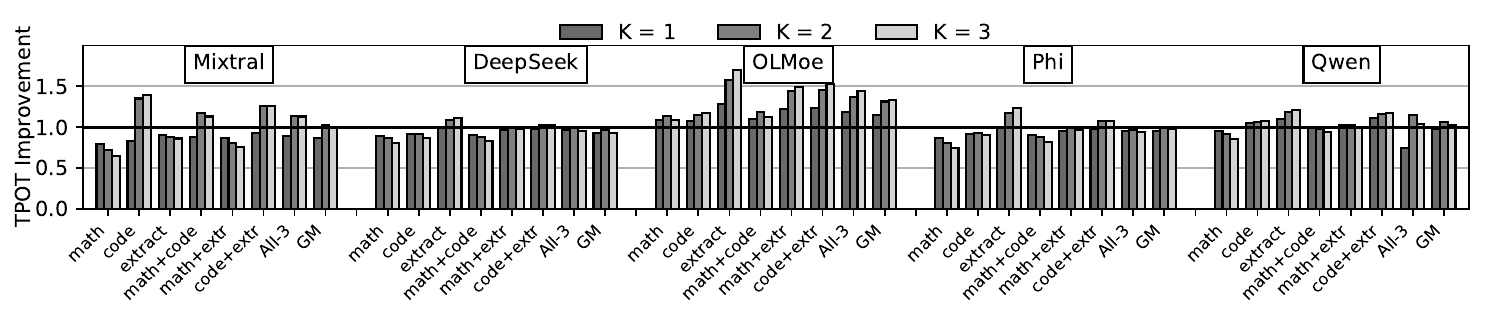}
    \vspace{-2mm}

    \caption{TPOT improvement across popular MoE models for different same-request and mixed-request tasks at varying speculation lengths. For a given model, the same $K$-value provides speedup in some tasks while incurring performance loss in others, making selection of $K$ difficult. Overall, speculation largely remains ineffective for most MoEs.}
    \label{fig:perf_variation_K}
    \vspace{-2mm}
\end{figure*}

\subsection{Speculation Dependence on Model and Task}


Effective speculation in MoEs is challenging because the costs and benefits of speculation and the resulting TPOT speedup vary with model architecture and request stream. To illustrate, \Cref{fig:perf_variation_K} shows TPOT improvements for $K \in {1, 2, 3}$ across five popular MoE models and diverse workloads. We note that higher $K$ values do not appreciably improve performance due to rising verification costs. In addition to standard tasks like code and math, we evaluate mixed workloads, where a single model concurrently serves heterogeneous requests (e.g., code+math), a common scenario in real-world deployments. Evaluation methodology details are available in \cref{sec:exp_setup}.

We make two key observations.
\textbf{First}, no speculation length improves performance across all tasks for any model we evaluated. For example, Mixtral at $K{=}3$ delivers a 39\% speedup on code but incurs a 54\% slowdown on math. When serving a mixed workload (50\% math, 50\% code), the performance gain drops to 13\%. This variability renders static-$K$ profiling ineffective in real-world serving scenarios with heterogeneous requests.
\textbf{Second}, disabling speculation ($K{=}0$) is optimal for some model–task pairs. Phi MoE~\cite{abdin2024phi}, for instance, incurs 10\% slowdown on math at $K{=}1$. Although extraction tasks benefit from up to 20\% speedup at $K{=}3$, a math+extraction mix sees no improvement at any static $K$. As a result, even adaptive-$K$ schemes must include \textit{no speculation} as a viable choice.

\vspace{-4mm}

\subsection{Limitations of Prior Works}

To maximize speculation efficacy and avoid slowdown, systems must adapt the $K$-value based on incoming requests. Prior work on dense models~\cite{brown2024ddd, mamou2024disco, zhang2024svip} improves TPOT by dynamically tuning $K$, using the drafter's output probabilities to guide their decisions. However, these approaches are incompatible with model-free drafters like n-gram speculation, which lack the drafter's probability distribution. More critically, such schemes maximize ETR and cannot anticipate when \textit{no speculation} ($K{=}0$) is optimal. Even if they decide to stop speculating after one draft token, we would not know whether to verify the generated draft token. 
In MoEs, even a conservative ($K{=}1$) can introduce unacceptable overheads; for example, Mixtral incurs a 25\% slowdown on the math task (\cref{fig:perf_variation_K}). Thus, current dynamic-$K$ policies are infeasible for MoEs.

\subsection{Variation of ETR and Speculation Cost}
\label{sec:etr_cost_variation}

\begin{figure}[htb!]
\centering
\subfloat{%
    \includegraphics[width=0.8
    \columnwidth]{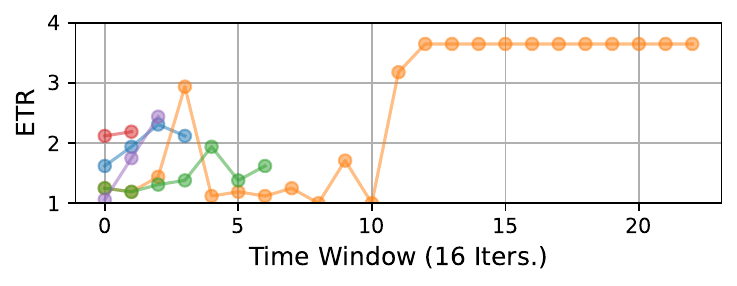}
}
\vspace{-8mm}

\subfloat{%
    \includegraphics[width=0.8\columnwidth]{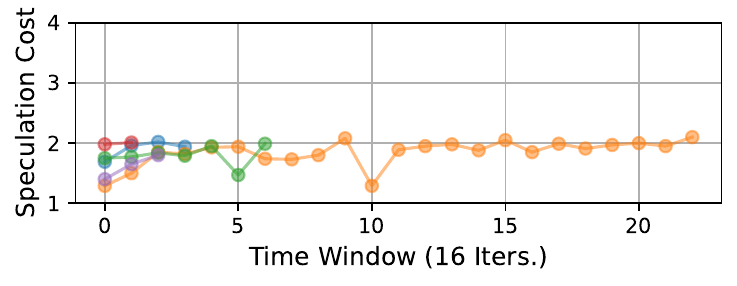}
}
\vspace{-3mm}
\caption{Iteration-level variation of ETR and speculation cost for Phi MoE serving extraction requests (23\% TPOT speedup).}
\label{fig:phi_mtbe}
\end{figure}

To navigate the cost-benefit tradeoff of speculation at runtime, we observe that both ETR and speculation cost, while variable across tasks, requests, and iterations, exhibit temporal locality over short iteration intervals. To illustrate, \Cref{fig:phi_mtbe} plots ETR and cost variation (averaged over 16 iterations) for five extraction queries served by the Phi MoE at static $K{=}3$. Despite occasional phase shifts, recent ETR and cost information reliably predict near-future trends. For example, beyond the $10^{\text{th}}$ window on the yellow curve, the ETR gain exceeds the cost, making speculation effective.

Our design exploits the temporal locality of ETR and speculation cost to adjust $K$ dynamically at the iteration level. We introduce a unified metric, the \textit{utility} of speculation, that captures both the cost and the benefit of speculation. We show in \cref{sec:utility} that utility reliably predicts speculation performance, and maximizing it directly minimizes TPOT, making it suitable to guide $K$-value selection.


\vspace{-2mm}

\subsection{Goal of the Paper}

The goal of this work is to maximize the efficacy of speculation for low-latency MoE inference in real-world serving scenarios. To do so, the solution must adapt to heterogeneous request streams and generalize across speculation techniques, MoE architectures, and hardware platforms. To this end, we introduce a utility-driven speculation management framework that dynamically tunes $K$ and disables speculation when it is detrimental, making speculation both practical and effective for MoE serving.

\mqignore{
\begin{figure}[h]
    \centering    
           \includegraphics[width=.8\columnwidth]{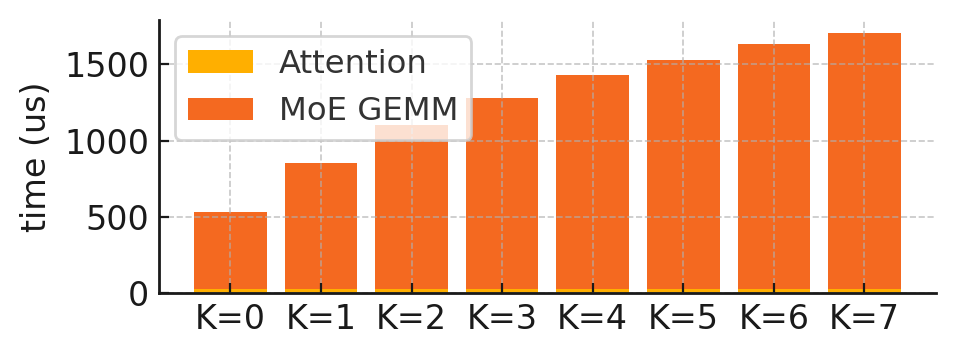}
    \caption{\mytodo{Update with real data.} Figure for Sec 4.2.}
    \label{fig:iter_time_breakdown}
    \vspace{-3mm}
\end{figure}

\smallskip
\noindent\textbf{Breakdown of Verification Step.} At each verification iteration, $K+1$ in-flight tokens (for speculation length of $K$) are passed through the entire MoE, comprising of transformer blocks with attention and expert FFN layers. In \cref{fig:iter_time_breakdown}, we plot the runtime breakdown of time taken by the attention and expert FFN layers in the Mixtral MoE model at different speculation lengths for N-Gram speculation. Attention takes up a small fraction of the overall latency (8\% for baseline) and its computation time remains the same across all speculation lengths. The expert FFN layer, however, takes up the majority of the latency and its computation time increases with speculation length, which we investigate next.
}

\newpage

\section{Experimental Setup}
\label{sec:exp_setup}

\smallskip
\noindent\textbf{Serving Framework.} We evaluate on the vLLM serving framework and implement our solution using Python and PyTorch operators.

\begin{table}[h]
    \scriptsize
    \vspace{-0.1 in}
    \caption{MoE models evaluated in this work.}
        \vspace{-0.05 in}
    \begin{tabular}{|c|c|c|c|c|c|}
    \hline
    \textbf{\begin{tabular}[c]{@{}c@{}}MoE\\ Model\end{tabular}} & \textbf{\begin{tabular}[c]{@{}c@{}}Hidden,\\ Layers\end{tabular}} & \textbf{\begin{tabular}[c]{@{}c@{}}Active/\\ Total P\end{tabular}} & \textbf{\begin{tabular}[c]{@{}c@{}}Total\\ Exps.\end{tabular}} & \textbf{\begin{tabular}[c]{@{}c@{}}Top-K\\ Exps.\end{tabular}} & \textbf{\begin{tabular}[c]{@{}c@{}}Shared\\ Exps.\end{tabular}} \\ \hline
    \textbf{Mixtral FP8}                                             & 4K, 32                                                            & 13B/ 47B                                                           & 8                                                              & 2                                                              & 0                                                               \\ \hline
    \textbf{Phi-3.5 FP8}                                             & 4K, 32                                                            & 6.6B/ 42B                                                          & 16                                                             & 2                                                              & 0                                                               \\ \hline
    \textbf{OLMoE FP8}                                               & 2K, 16                                                            & 1B/ 7B                                                             & 64                                                             & 8                                                              & 0                                                               \\ \hline
    \textbf{DeepSeekV1 FP16}                                         & 2K, 28                                                            & 2.8B/ 16.4B                                                        & 66                                                             & 6                                                              & 2                                                               \\ \hline
    \textbf{Qwen-1.5 FP16}                                            & 2K, 24                                                            & 2.7B/ 14B                                                          & 64                                                             & 4                                                              & 4                                                               \\ \hline
    \end{tabular}
    \vspace{-1mm}
    \label{tab:models}
    \end{table}

\smallskip
\noindent\textbf{Models.} We evaluate five popular MoEs: Mixtral, DeepSeek-v1, Qwen-1.5, Phi-3.5, and OLMoE~\cite{jiang2024mixtral, deepseekv1, qwen_moe, abdin2024phi, muennighoff2024olmoe}. We note that our models cover a variety of MoE architectures, including shared-experts employed in DeepSeek and Qwen, as well as quantized models (FP8). The details of the models are provided in Table \ref{tab:models}

\smallskip
\noindent\textbf{Workloads.} We evaluate on popular decode-heavy tasks: mathematical reasoning, code generation, and multi-turn conversation using GSM8K, HumanEval, and MT-Bench datasets~\cite{gsm8k,humaneval,mtbench}, respectively. For MT-Bench, we use the natural language "extraction" task. We also develop four mixed workloads that comprise request streams from 2 or 3 tasks with equal sharing of requests. For instance, the code+math task has 50\% share of code and math requests each, while the All-3 task has 33\% requests from code, math, and extraction. All mixed workloads run for 10 minutes (comparable to same-request tasks) and generate at least 20,000 tokens.


\smallskip
\noindent\textbf{Speculative Decoding Techniques.} 
We evaluate all MoE models with n-gram speculation~\cite{fu2023lookahead,saxena2023prompt}. We also evaluate the state-of-the-art draft-model based EAGLE~\cite{li2024eagle} technique on Mixtral, as pretrained drafters for EAGLE are  not available for other MoEs. We use a maximum $K$ of 3 for all our static-K schemes as increasing K further yields, at-best, marginal benefits.

\smallskip
\noindent\textbf{Hardware.}
We conduct experiments on a workstation employing an RTX 6000 Ada GPU with 48GB memory and a 24-core (32-thread) Intel i9-14900K CPU with 64GB memory. 

\smallskip
\noindent\textbf{Baselines.} For each model and speculation technique, we provide results for static speculation-lengths between 1 and 3. In our graphs, no-speculation is implicitly plotted as the y=1 line.

\smallskip
\noindent\textbf{Figures of Merit.} We measure the output token generation throughput (toks/sec), which is the inverse of the time per output token (TPOT) in our single-batch serving setup. We also measure the effective-token rate and overheads of speculation techniques.

\begin{figure*}
    \centering    
           \includegraphics[width=1.95\columnwidth]{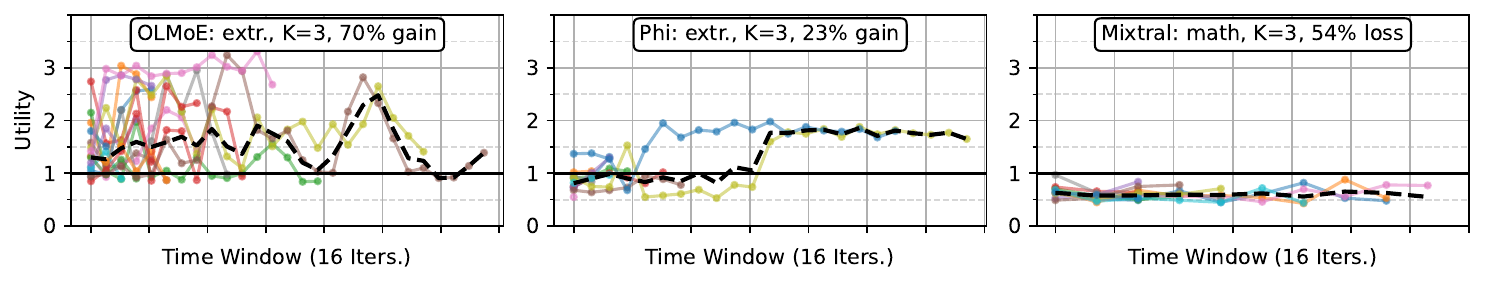}
    \vspace{-2mm}

    \caption{Utility variation for different requests, averaged for window of 16 iterations, for selected model, task, and speculation length combinations. Each plot-line is a request, and the dotted black line is the moving harmonic mean utility across requests.  Utility varies from request to request, and also across iterations for the same request.}
    \label{fig:utility_iter_variation}
    \vspace{-2mm}
\end{figure*}


\newpage


\section{Speculation Utility}
\label{sec:utility}

Maximizing speculation efficacy requires balancing its costs and benefits when tuning the $K$-value. To this end, we develop a simple heuristic—\textit{speculation utility}—that guides our speculation management policy. 
We define the speculation utility below.

\begin{definition}[Speculation Utility]
\label{def:utility}
The ratio of the benefit of speculation (ETR improvement) to its cost (speculation overheads). 
\end{definition}

\begin{figure}[htb]
    \centering    
           \includegraphics[width=.9\columnwidth]{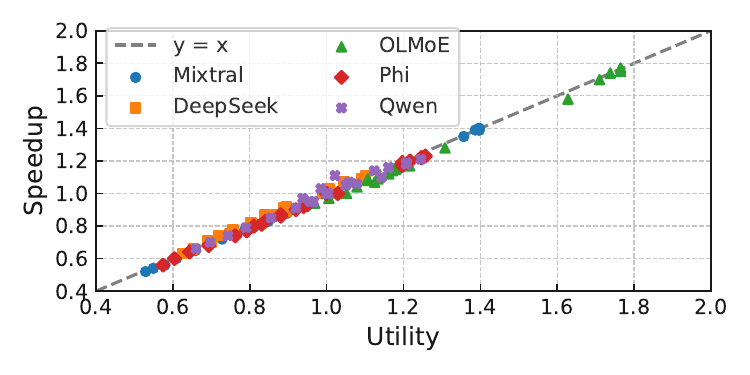}
    \vspace{-2mm}
    \caption{Speedup as a function of measured utility for 5 MoEs, 3 tasks, and 8 static speculation-length combinations. Utility reliably predicts performance ($R^2=99.4\%$).}
    \label{fig:utility_speedup}
    \vspace{-2mm}
\end{figure}

To illustrate the effectiveness of utility, \Cref{fig:utility_speedup} plots average utility values across 120 datapoints, spanning 5 MoE model architectures, 3 tasks (code, math, and extraction), and 8 static speculation settings ($K \in [0, 7]$) using n-gram speculation. In our setup, $K=0$ denotes no speculation. We compute speculation cost as the ratio of the average iteration time under speculation, including drafting, verification, and rejection sampling, to the baseline iteration time without speculation.
This analysis yields three key insights.
\textbf{First}, unlike ETR, \textit{utility robustly predicts speculation performance} by accounting for both cost and benefit.
\textbf{Second}, when utility falls below one, speculation no longer yields net gains, making utility an effective criterion for \textit{dynamically disabling speculation} at runtime.
\textbf{Third}, \textit{maximizing utility directly maximizes TPOT speedup}. We formalize this relationship in the following theorem.


\begin{theorem}
\label{thm:utility}
If, on average, the time per output time (TPOT) of a non-speculative baseline is $t_{base}$, and the TPOT of speculation with utility of $U_{spec}$ is $t_{spec}$, then $t_{spec} = t_{base}/U_{spec}$.
\end{theorem}


\smallskip
\noindent\textbf{Proof.} TPOT (time per output token) is defined as the inverse of tokens per second (TPS) throughput:
\begin{equation}
TPOT = \frac{1}{TPS}, \quad TPS = \frac{toks}{iter}\cdot\frac{iters}{sec} = \frac{ETR}{t_{iter}}
\end{equation}
Here, $ETR$ is the effective token rate, and $t_{iter}$ is iteration execution time. For the baseline (no speculation), $ETR_{base}=1$.
We defined speculation's \textit{cost} as the ratio of speculative to baseline iteration times, and its \textit{benefit} as the speculative ETR:
\begin{equation}
\text{cost} = \frac{t_{iter,spec}}{t_{iter,base}}, \quad \text{benefit} = ETR_{spec}
\end{equation}

Utility ($U_{spec}$) thus becomes:
\begin{equation}
U_{spec} = \frac{ETR_{spec}}{t_{iter,spec}/t_{iter,base}}
\end{equation}

Noting $t_{base}=t_{iter,base}$ and $t_{spec}=t_{iter,spec}/ETR_{spec}$, the ratio of baseline to speculative time simplifies to:
\begin{equation}
\frac{t_{base}}{t_{spec}} = \frac{ETR_{spec}}{t_{iter,spec}/t_{iter,base}} = U_{spec}
\end{equation}

Rearranging yields $t_{spec} = t_{base}/U_{spec}$.

\smallskip
\noindent
Therefore, by maximizing iteration-level utility, we can minimize end-to-end TPOT with speculation, as desired. However, like TPOT improvement, utility is not trivial to predict, due to variation between models and tasks, and even across requests and iterations.

\subsection{Speculation Utility Characterization}
\label{sec:utility_characteristics}

\Cref{fig:perf_variation_K} shows significant variation in TPOT speedup (and by extension, utility) across MoE architectures, tasks, and speculation lengths. Static profiling to select optimal $k$-values cannot capture dynamic request patterns.
Therefore, we must compute utility at runtime at request-level and iteration-level granularity.

In \Cref{fig:utility_iter_variation}, we show that utility varies significantly across requests, even within the same task. The figure plots utility averaged over sliding windows of 16 iterations for various model–task–$K$ combinations. Some combinations exhibit higher utility later in generation (e.g., Phi with extraction), while others maintain consistently low utility (e.g., Mixtral with math). Individual requests also exhibit phases of high and low utility, such as OLMoE with extraction.
Despite these variations, utility remains relatively stable over short sequences of intervals, reflecting the \textit{temporal locality} of ETR and speculation cost trends (\cref{sec:etr_cost_variation}). Declining utility often signals the need to reduce $K$ to avoid overheads. Conversely, requests with low initial utility may become more favorable deeper into generation. To capture the dynamic behavior and maximize speculation benefits, we monitor utility at iteration-level granularity.


\smallskip
\begin{tcolorbox}[boxrule=1pt,left=5pt,right=5pt,top=1.5pt,bottom=1.5pt]
    \mytodo{variability across what? middle ground vs low level vs high level}
    \textbf{Takeaway:} \textit{Despite utility variation in a request's lifetime, we observe temporal locality within short intervals, and we use recent utility information to make reliable predictions for the near future.} 
\end{tcolorbox}


\section{Utility-Driven Speculative Decoding}
\label{sec:cost_aware_sd} 


\subsection{Design Overview}
\label{sec:casd_overview}

\begin{figure}[h]
    \centering    
           \includegraphics[width=.975\columnwidth]{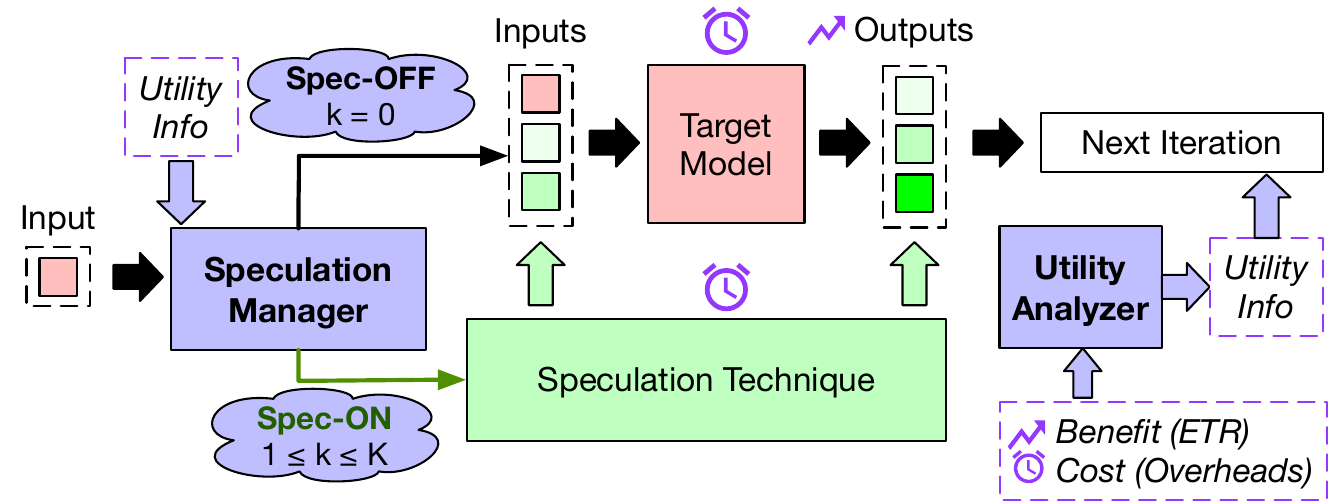}
    \caption{Overview of utility-driven speculation. The speculation manager decides whether to enable speculation, and if so, the spec-length $k$. The utility analyzer keeps track of recent costs and benefits of speculation to aid the manager.
    }
    \label{fig:casd_overview}
    \vspace{-2mm}
\end{figure}

To manage speculation trade-offs in MoEs, we propose \TheName, a cost-aware speculative decoding framework that dynamically adjusts the speculation length ($K$) based on the current \textit{utility}, which is the ratio of speculation’s benefit (ETR) to its cost (verification overheads). Beyond being a robust predictor of speculation performance, utility enables \TheName to operate independently of model architecture, task, hardware, or speculation method. \Cref{fig:casd_overview} illustrates \TheName's design, which consists of two components: a \textit{speculation manager} that controls speculation settings (e.g., enabling/disabling speculation and tuning $K$), and a \textit{utility analyzer} that tracks costs and benefits to compute utility information at runtime.


Decisions in \TheName are utility-driven. If predicted utility falls below 1, the speculation manager disables speculation to avoid slowdown. When utility exceeds 1, the manager chooses the speculation length that maximizes utility. For simplicity, we denote disabled speculation as $k=0$, and positive $k$ values imply speculation is enabled and indicate the speculation length. 

\subsection{Design Considerations}
\label{sec:casd_considerations}


In subsequent sections, we address the following key questions that inform our design choices in \TheName.

\begin{enumerate}
    \item \textbf{Preventing Slowdown.} How to quickly determine if speculation is beneficial? It utility is below 1, testing a $k$-value greater than 0 may incur significant overheads.
    \item \textbf{Speculation Tuning Granularity.} How often should the speculation manager decide the $k$-value? It can range from deciding every iteration to setting $k$-value for the duration of a request, or even across requests.
    \item \textbf{Utility Information Granularity.} What utility information should the utility analyzer provide to the manager? The utility information can be provided at varying granularity, such as offline profiled per-task utility, or utility measured online, as often as every iteration.
    \item \textbf{Maximizing Utility.} How would the speculation manager arrive at the $k$-value that maximizes utility?
\end{enumerate}



\subsection{Leveraging Utility: Test-and-Set Policy}
\label{sec:test_and_set_policy}

\mqignore{
\smallskip
\noindent\textit{Test-and-Set to Maximize Utility.} We find that utility, while dynamic, exhibits locality across small intervals of iterations, allowing us to periodically make decisions using a \textit{test-and-set} policy.
In the test-phase, we try out up-to $M$ distinct $K$-values in quick succession of $T$ iterations, while keeping track of associated utility.
At the end of the test-phase, we set the $K$-value that maximizes the utility for the upcoming set phase, which comprises the next $S$ iterations.
The test phase is small, comprising up-to four trials ($M$) of four iterations each ($T$).
Similarly, the set phase is kept at 16 iterations ($S$), to enable adapting to dynamic changes in $K$-value.
If utility during the test phase is below 1 even for $K=1$, we disable speculation for the set phase, minimizing performance loss.

\smallskip
\noindent\textit{Adaptive Back-off to Minimize Test Cost.} Even if speculation is disabled, its cost can still be significant. For instance, if current utility is 0.5 (or 2x slowdown), then spending even 4 out of 20 iterations ($T+S$) in speculative state would incur 20\% slowdown. 
To reduce slowdown when utility is consistently below 1, we minimize the cost of testing by increasing the set-phase duration by 2x~(with $K=0$) every time we drop again to this state. During testing phase, we try out $K$ values as usual, ensuring we still capture opportunities to speculate effectively. 
Adaptively backing-off reduces the worst-case slowdown encountered by \TheName from 15\% to 5\%. 

\smallskip
\noindent\textit{Hill-Climbing to Find Utility-Maximizing $K$.} As utility exhibits locality across iteration-intervals, the \textit{direction} of change in utility provides clues to choosing the best-K.
For instance, if utility decreases by increasing $K$, then costs of verifying more tokens is likely dominating the benefits afforded by improved ETR, and we should back-track to a lower $K$.
Our observation on directionality of utility inspire our hill-climbing search to tune $K$ dynamically.
We find that a few trials of $K$-values are sufficient to establish \textit{best-K} for the next interval.
While profiling static-K values for every request stream is infeasible for practical deployment, \TheName performs comparably to such best-performing static-K schemes for all evaluated tasks, with a maximum average gain of 14\% across tasks for Mixtral. 
}

\begin{figure}
    \centering    
           \includegraphics[width=.9\columnwidth]{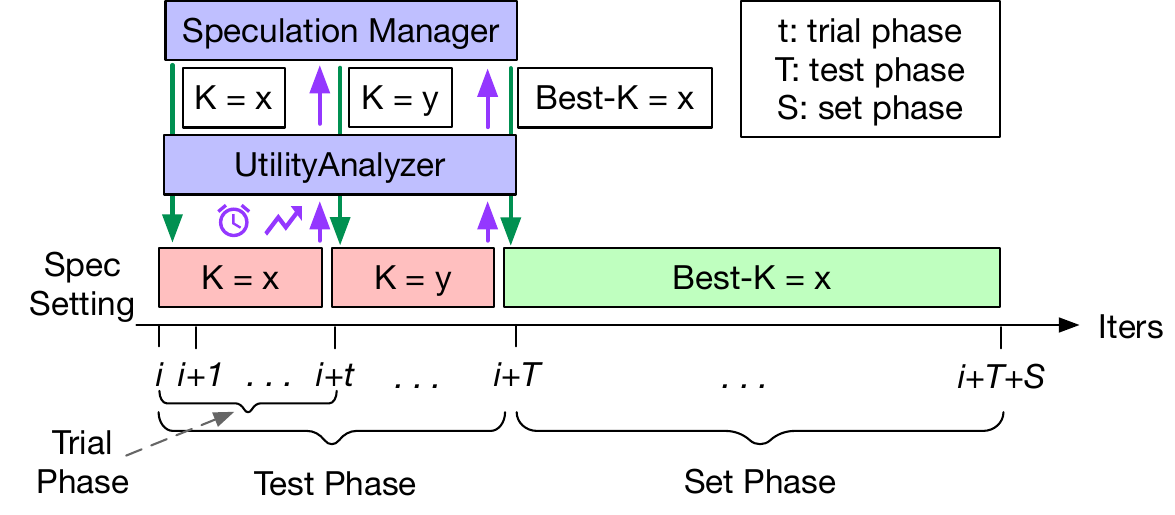}
    \caption{Test-and-Set Policy: Periodically, the speculation manager perform trials of different K-values in the test phase, and uses the best-performing $K$ for the set phase.}
    \label{fig:test_and_set}
    \vspace{-2mm}
\end{figure}

Our utility characterization shows that while utility varies dynamically, it exhibits locality over short \textit{intervals} of iterations. Therefore, we make speculation decisions periodically, rather than at every iteration. At the end of each interval, we compute the utility of recent iterations and use it to select the $K$-value for the next interval. However, committing to a $K$-value without verification risks high overhead if the prediction is inaccurate.
To address this, we propose a \textit{test-and-set} policy: the speculation manager briefly tests different $K$-values over a few iterations, measures their utility, and then sets the $K$-value for the upcoming interval based on their utility. Our policy adapts to utility variation rapidly and selects $K$ efficiently by exploring multiple values in quick succession.

\Cref{fig:test_and_set} illustrates our test-and-set policy. During the \textit{test} phase, the speculation manager evaluates different $K$-values over short \textit{trial} periods of $t$ iterations each (e.g., $t=4$). To bound testing overhead, we limit the number of trials to a small value, $M$ (e.g., $M=4$), particularly when speculation is unprofitable. We describe our method for selecting the utility-maximizing $K$ in \cref{sec:dynamic_spec_length}. After at most $T = M \cdot t$ test iterations, the manager selects the best-performing $K$ and enters the \textit{set} phase, where it uses this value for the next $S$ iterations.
The first trial uses a starting value, $K_{\text{start}}$, which the manager chooses dynamically by scanning recent history and selecting the non-zero $K$ that previously yielded the highest utility. To compute speculation cost, the utility analyzer normalizes iteration time to a no-speculation baseline. We obtain the baseline iteration time efficiently by running the first few decode iterations without speculation (e.g., 4) and refreshing the no-speculation measurement infrequently (e.g., every 100 iterations).


\subsection{Preventing Slowdown: Disable Speculation}
\label{sec:prevent_slowdown}

Speculation in MoEs can incur prohibitive overhead when costs outweigh benefits. To avoid slowdowns, we compute utility at the end of each trial phase and disable speculation by setting $K{=}0$ for the next set phase if utility falls below one. If the current $K$ is already 1, we exit the test phase early to avoid further unnecessary overhead.
In speculative decoding, rejection sampler accepts tokens \textit{causally}—i.e., the acceptance of later tokens depends on the acceptance of earlier ones. As a result, $K{=}1$ represents the most conservative speculative state, with the lowest possible overhead.
We continue monitoring utility by restarting the test phase after every $t{+}S$ iterations, beginning each cycle with $K{=}1$. This approach allows the framework to detect utility changes during longer sequences and re-enable speculation if it becomes beneficial.

\subsection{Minimizing Test Cost: Adaptive Back-Off}
\label{sec:adaptive_back_off}

As discussed in \cref{sec:utility_characteristics}, some requests benefit from speculation later in generation—often because the drafter produces more accurate tokens with additional context. To capture such delayed benefits, we re-evaluate utility every $t{+}S$ iterations, as detailed in \cref{sec:prevent_slowdown}.
However, for requests where speculation consistently underperforms (i.e., utility remains below 1), repeated testing can introduce significant overhead. Consider $t{=}4$ and $S{=}16$. In models like Mixtral and Phi, speculative decoding can incur up to 3x the cost of no-speculation, especially when testing begins with a high $K_{\text{start}}$. Even at 2x overhead and zero benefit (ETR = 1), the test phase alone takes 8 time units, and the full 20-iteration trial+set interval consumes 24 units, 20\% more than the no-speculation baseline. This overhead occurs despite disabling speculation in the set phase.

\begin{figure}[htb!]
    \centering    
           \includegraphics[width=.9\columnwidth]{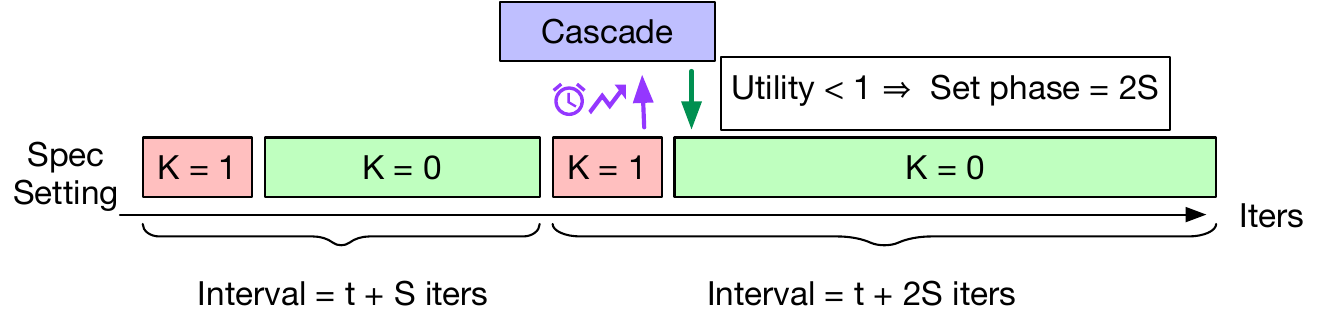}
    \caption{Adaptive Back-Off: If utility at $K=1$ is persistently below 1, test less frequently to minimize cost.}
    \label{fig:adaptive_back_off}
    \vspace{-3mm}
\end{figure}

To remain practical, \TheName must impose minimal overhead even when speculation consistently fails to provide benefits. To achieve this, we implement an \textit{adaptive back-off} strategy. Since $K_{\text{start}}$ is always non-zero, we treat any transition to $K=0$ during a sequence as an indicator to reduce testing frequency. Specifically, we double the set phase length $S$ whenever $K$ drops to 0. \Cref{fig:adaptive_back_off} provides an overview of this optimization.
Adaptive back-off minimizes the cost of repeated testing by reducing its frequency exponentially. In the earlier example, over 250 iterations, \TheName would run speculative trials for only 18 iterations compared to nearly 50 without back-off. As a result, the worst-case slowdown in scenarios where speculation is harmful (e.g., Mixtral + math) is just 5\%, versus 14\% without adaptive back-off, as we evaluate in \cref{sec:midas_opts_sens}.


\subsection{Maximizing Utility: Hill-Climbing Search}
\label{sec:dynamic_spec_length}

\begin{figure}[h]
    \centering    
           \includegraphics[width=.6\columnwidth]{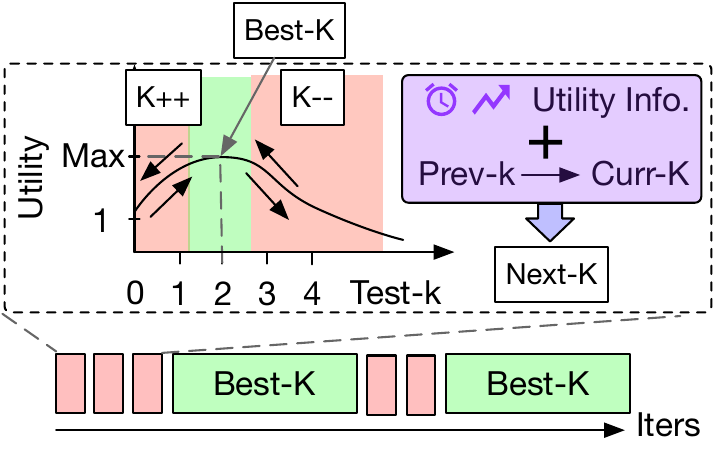}
    \caption{Hill climbing search: Leverage the directionality of utility with respect to K to find best-K.}
    \label{fig:hill_climbing_search}
    \vspace{-5mm}
\end{figure}

Since utility exhibits temporal locality across iteration intervals, its \textit{direction of change} provides clues for identifying the best $K$. \TheName leverages this by performing a hill-climbing search during the test phase to select the optimal $K$ for the subsequent set phase. \Cref{fig:hill_climbing_search} illustrates the search procedure. The inputs to the search are the utility values and corresponding $K$-values from the current and previous trials (denoted as \texttt{curr-K} and \texttt{prev-K} in the figure). By comparing these values, the speculation manager infers the direction in which utility increases to determine \texttt{next-K}. The manager continues searching for $M$ trials to converge on a utility-maximizing $K$.
The intuition behind this search is based on our observation that utility varies directionally with $K$. As $K$ increases, utility improves up to a point, beyond which costs begin to outweigh benefits.


\begin{figure*}
    \centering    
           \includegraphics[width=1.95\columnwidth]{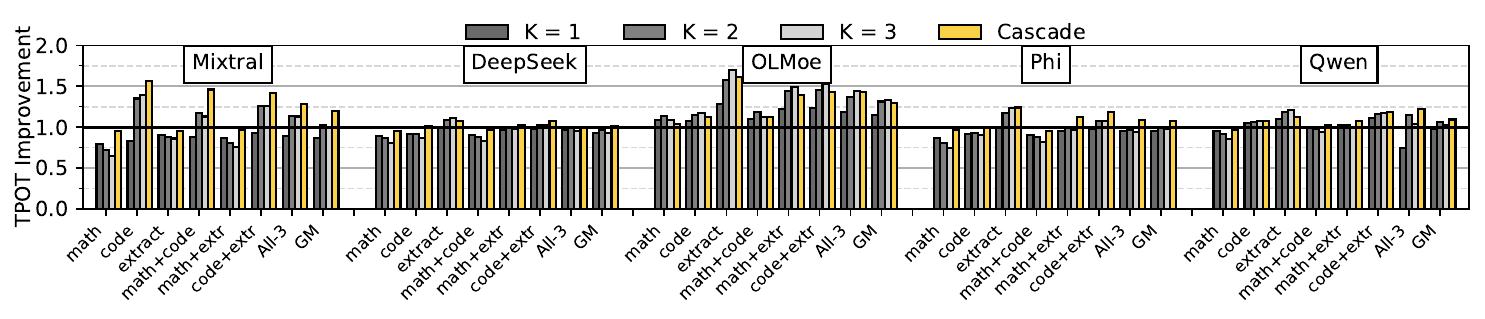}    \caption{TPOT improvement of \TheName and static-K schemes on 5 MoE and 7 tasks with n-gram speculation. \TheName outperforms static-K schemes for all models except OLMoE (3\% average slowdown compared to static K=3). On average, \TheName improves performance over static-K schemes by 7-15\%, and by up-to 1.6x compared to a non-speculative baseline. }
    \label{fig:perf_results}
    \vspace{-2mm}
\end{figure*}

For example, if the previous $K$ is lower than the current $K$, and utility increases, then further increasing $K$ may improve utility. Conversely, if utility decreases with a higher $K$, the cost of verifying additional tokens likely outweighs the ETR gains, and the manager should backtrack to a lower $K$. Since the utility landscape is not known \textit{a priori}, \TheName tracks the $K$-value that yields the highest utility during the test phase. At the end of testing, the manager selects this best-performing $K$ (denoted \texttt{best-K}) for the set phase.
To reduce unnecessary trials, we exit the test phase early under three conditions: (1) utility consistently decreases across successive trials, indicating a local maximum has been passed; (2) the current $K$ reaches 0, signaling speculation is no longer beneficial; or (3) utility values across successive $K$-values converge within 10\%, suggesting diminishing returns and proximity to the optimum.


\subsection{Generalizability of \TheName}

\TheName is a utility-driven framework that computes utility as a simple ratio of benefit (ETR) to cost (iteration time overhead) under speculation. Its principled design makes it broadly applicable across speculation techniques, model architectures, heterogeneous tasks, and hardware platforms. To demonstrate its versatility, we present case studies using EAGLE drafter-based speculation in \cref{sec:eagle}.
Unlike prior speculation systems, \TheName requires no offline profiling or detailed performance modeling to guide decisions. Additionally, unlike drafter-centric dynamic speculation techniques that rely on trained neural networks to predict $K$, our hill-climbing search is lightweight, intuitive, and agnostic to both task and model. Improved $K$-search strategies can be seamlessly integrated, and techniques that reduce MoE-specific speculation overheads complement \TheName. We discuss several such optimizations in \cref{sec:related}.
Finally, although we implement \TheName in the vLLM framework, the design is readily portable to other production systems such as NVIDIA’s TensorRT-LLM~\cite{trtllm} and HuggingFace’s TGI~\cite{tgi}.

\section{\TheName Implementation}

\begin{figure}[htb!]
    \centering    
           \includegraphics[width=.9\columnwidth]{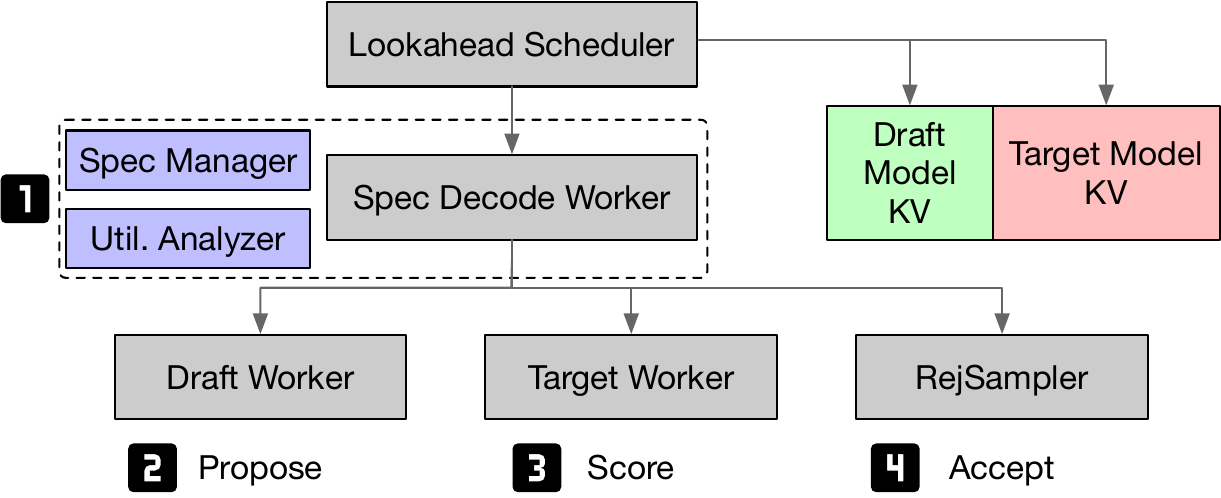}
    \caption{System architecture of speculation in vLLM. We modify the spec-decode worker class to implement \TheName.}
    \label{fig:spec_decode_vllm}
    \vspace{-3mm}
\end{figure}

We implement \TheName on vLLM serving framework (v0.6.6), using approximately 500 LoC (Python/ PyTorch). The vLLM runtime manages KV-cache states through a memory manager and handles incoming request scheduling using a lookahead scheduler, which reserves speculative generated token KV-states. The running requests are dispatched to the \textit{spec-decode worker}, which is responsible for coordinating the speculation workflow. As illustrated in \Cref{fig:spec_decode_vllm}, we integrate our modifications into the speculation-decode worker, which runs on the CPU. This worker calls the draft and target models and performs rejection sampling at every decoding iteration.


\newpage
To implement \TheName, we make the following key changes:

\begin{itemize} 
    \item \textbf{Dynamic Speculation Support:} vLLM does not support re-enabling speculation once it is disabled for a request. We implemented dynamic speculation-disable by maintaining consistent KV-cache states by executing the drafter even when speculation is off, incurring a modest overhead of 2–3\% on average. While n-gram does not have KV-cache states, our changes target model-based speculation techniques, such as EAGLE, which we also evaluate. 
    \item \textbf{Utility Analysis Telemetry:} vLLM already tracks iteration-level scoring, drafting, and proposal times. We implemented lightweight utility computation kernels that run on the CPU, and incur negligible runtime relative to an MoE iteration of between 6ms (OLMoE) and 28ms (Mixtral). 
    \item \textbf{Speculation Manager:} We modified the speculation decode worker's execute-model functionality to handle no-speculation and adaptive speculation length dynamically. This management logic, consisting of simple conditional statements and arithmetic operations, and runs on the CPU.
\end{itemize}

\smallskip
\noindent\textbf{Hyperparameters.}
\TheName does not rely on model, task, or hardware-specific hyperparameters, contributing to its generalizability across diverse deployment settings. The only required parameters pertain to the test-and-set policy: $t$ (trial phase duration), $T$ (maximum test phase length), and $S$ (set phase duration). To minimize overhead, the set phase should be significantly longer than the test phase, while keeping the test phase short allows rapid iteration over candidate $K$-values. In our evaluations, we find that $t=4$, $T=16$ (up to 4 trials), and $S=16$ provide a practical balance between responsiveness to changing utility and efficient testing. We analyze the sensitivity of our choices in \cref{sec:hyperparam}.


\section{Evaluation Results}
\mytodo{Bullet points for evals.}

\Cref{fig:perf_results} presents the performance of \TheName across five MoE models and seven tasks, compared to static-$K$ baselines with $K \in {1, 2, 3}$ ($y{=}1$ denotes no speculation). Static schemes suffer from high worst-case slowdowns of up to 26\%, 38\%, and 54\% for $K=1$, $2$, and $3$, respectively. In contrast, \TheName limits worst-case slowdown to just 5\% and improves average performance across all MoEs except DeepSeek, which sees a marginal 1\% gain.
On average, static schemes provide marginal benefit, except for OLMoE, where $K=2$ and $3$ yield 31\% and 33\% speedups, respectively, compared to 29\% with \TheName. This small drop is due to testing phases, where temporarily lower $K$ values slightly reduce performance. OLMoE's higher speculation gains arise from strong expert-to-token affinity~\cite{muennighoff2024olmoe}, which lowers verification cost and amplifies ETR benefits.


Among the evaluated models, Mixtral exhibits low expert-to-token affinity~\cite{muennighoff2024olmoe}, which increases the cost of speculation as more experts are activated. Thus, runtime adaptivity is critical for maintaining performance. \TheName achieves substantial gains on Mixtral, improving TPOT by 18\% over the best-performing static $K{=}2$ baseline, showing the effectiveness of dynamic speculation management.
Beyond preventing slowdowns (e.g., in the math task), dynamically tuning $K$ also improves performance in favorable scenarios. For instance, in the code task, \TheName achieves a 57\% speedup, compared to 39\% with a static $K{=}3$ policy. These results show that adaptivity benefits both cost-sensitive and speculation-friendly workloads.



\subsection{Case Study: Iteration-Level Utility Variation}

\begin{figure}[htb!]
    \centering
           \includegraphics[width=.9\columnwidth]{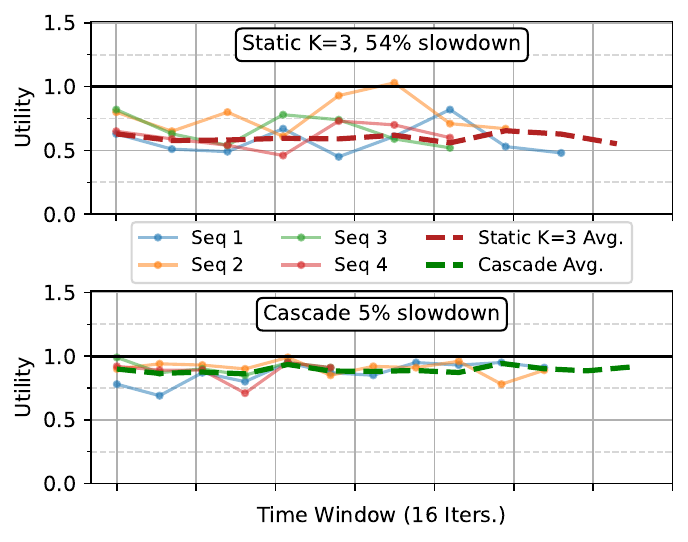}
        \vspace{-3mm}
    \caption{Utility variation in math+Mixtral using static K=3 and \TheName. Our scheme limits slowdown to 5\%.}
    \label{fig:util_comparison}
    \vspace{-3mm}
\end{figure}

A key feature of our design is the ability to disable speculation when utility falls below 1. However, many real-world deployments enforce strict service-level objectives (SLOs), requiring tight latency bounds per request and even on average TPOT. \Cref{fig:util_comparison} shows utility variation across iterations for four randomly selected requests, along with the task-level average (measured as the harmonic mean).
Static-$K$ schemes not only underperform but also exhibit high variability within requests, including instances with up to 2x slowdown. In contrast, \TheName maintains tighter TPOT bounds throughout generation, entering regions of high performance loss only rarely, most notably during test phases. Among the sampled requests, the maximum loss was 33\% and occurred only in 3 instances.
Overall, \TheName effectively limits degradation when utility is low, while preserving performance consistency across requests and tasks.


\subsection{Case Study: Task-Level Utility Variation}

\begin{figure}[htb!]
    \centering
           \includegraphics[width=\columnwidth]{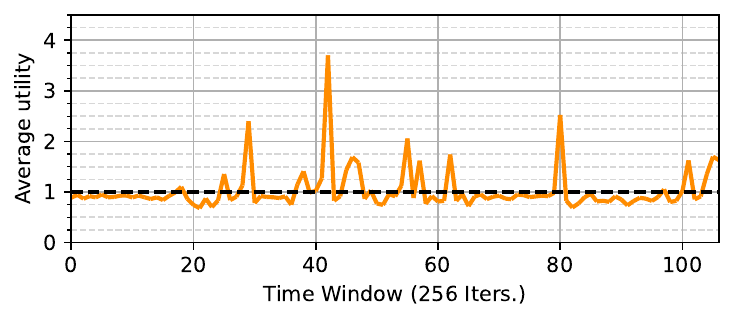}
        \vspace{-6mm}
    \caption{Utility variation for math+code+extraction mixed task with Mixtral over a 10-minute run with \TheName, which effectively adapts to request-level utility changes.}
    \label{fig:task_util_variation}
    \vspace{-4mm}
\end{figure}

For practical adoption, \TheName must maximize speculation efficacy across diverse requests. \Cref{fig:task_util_variation} shows runtime utility variation for the ALL-3 task using Mixtral, where the request stream consists of an even mix (33\% each) of code, math, and extraction queries. When speculation is not beneficial, \TheName minimizes slowdown by keeping utility near 1. When speculation is useful, it dynamically increases $K$ to maximize gains.
Overall, \TheName consistently maximizes utility across heterogeneous request streams.



\subsection{\TheName with EAGLE Speculation}
\label{sec:eagle}

\begin{figure}[htb!]
    \centering
           \includegraphics[width=\columnwidth]{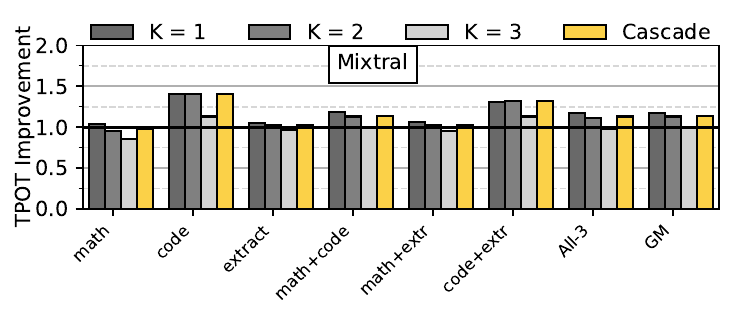}
        \vspace{-4mm}
    \caption{The performance of \TheName with EAGLE speculation. Our scheme performs comparably to static-K schemes. }
    \label{fig:eagle_spec}
    \vspace{-3mm}
\end{figure}

\TheName is a general-purpose framework compatible with any speculation strategy. We evaluate it using EAGLE, a state-of-the-art method that employs a feature-level autoregressive drafter to predict future token hidden states.
\footnote{We ported \TheName to vLLM v0.8.2 due to EAGLE-related bugs in v0.6.6.}
As EAGLE provides a pre-trained drafter only for Mixtral, we conduct all evaluations on that model. Figure~\ref{fig:eagle_spec} compares \TheName to static-$K$ baselines. On Mixtral, static-$K$ avoids slowdowns common with n-gram speculation due to EAGLE’s higher speculation accuracy (e.g., ETR of 1.7 vs. 1.3 on math at $K{=}1$). However, drafting overheads grow by ~5\% per unit increase in $K$, making $K=1$ the best-performing static setting. While \TheName yields only modest gains over static $K{=}1$, it consistently matches the best-performing static-$K$ across all evaluated tasks.

\subsection{Impact of \TheName Optimizations}
\label{sec:midas_opts_sens}

\begin{figure}[htb]
    \centering
           \includegraphics[width=\columnwidth]{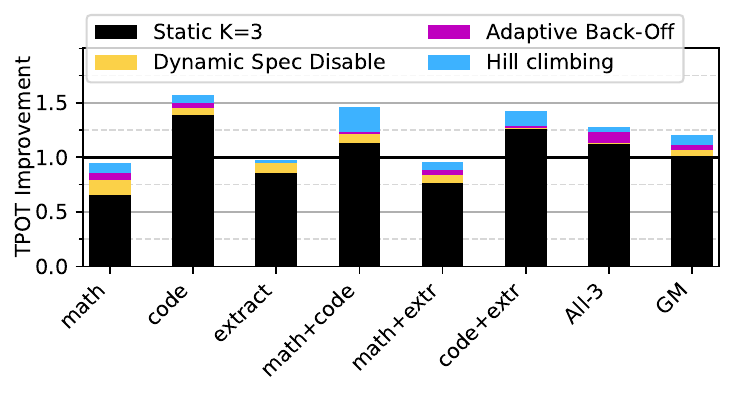}
    \vspace{-5mm}
    \caption{Impact of different optimizations of \TheName on performance of Mixtral. Our optimizations are additive.}
    \label{fig:optimization_sensitivity}
    \vspace{-3mm}
\end{figure}

\TheName incorporates three key optimizations: dynamic speculation disabling, adaptive back-off, and hill-climbing search. We isolate and incrementally enable each optimization and measure the resulting performance, as shown in Figure~\ref{fig:optimization_sensitivity}.
Without any optimizations, \TheName defaults to a static $K{=}3$ configuration (our chosen $k_{\text{start}}$).
Dynamic speculation disabling mitigates severe slowdowns when speculation offers little benefit. It also improves performance for tasks that benefit from speculation on average but include low-utility phases (e.g., code).
Adaptive back-off further reduces overhead on tasks that respond poorly to speculation (e.g., math).
Hill-climbing maximizes utility by adjusting $K$ based on observed trends, yielding gains across all workloads.
Overall, each of our optimizations meaningfully contribute to \TheName' effectiveness.


\subsection{Hyperparameter Sensitivity}
\label{sec:hyperparam}

\TheName exposes three hyperparameters that govern its test-and-set policy: trial ($t$), test ($T$), and set ($S$) phase durations. To support effective hill-climbing, we set $T = 4t$, enabling up to four trials per test phase. We vary $t \in {2, 4, 8}$ and set $S \in {8, 16, 32}$ accordingly, evaluating performance across all seven Mixtral tasks.
The small $t$-value of 2 yields noisy utility estimates due to high iteration-level variability, resulting in only a 13\% average gain over the no-speculation baseline. Conversely, large $S$-value of 32 misses fine-grained utility shifts and delays adaptation, limiting gains to 4\%.
These trends confirm that utility varies over time, yet exhibits short-term locality.
Our chosen configuration ($t{=}4$, $S{=}16$) balances responsiveness and reliability, achieving a 20\% average improvement.



\newpage
\section{Related Works}
\label{sec:related}

\subsection{Speculative Decoding Techniques}


\TheName is compatible with existing speculative decoding techniques.
Lookahead Decoding~\cite{fu2023lookahead} employs the Jacobi iteration method to parallelize autoregressive decoding without a draft model, generating $G$ multiple n-grams (each of length $K$) simultaneously.
This would significantly increase the in-flight tokens (by $G$) for MoEs, and our evaluations reveal that even vanilla n-gram decoding ($G=1$) incurs high costs. 
Medusa~\cite{cai2024medusa} integrates multiple decoding heads that predict several future tokens concurrently, with tree-based attention in target model to verify these predictions in parallel. 
Medusa also hurts MoE performance, as it increases the in-flight tokens by 50-100x and would activate all experts every iteration, for a cost increase of 4x-8x depending on the MoE sparsity, while the ETR increase rarely justifies the cost. 
The SoTA speculation technique, EAGLE, its successors~\cite{li2024eagle2,li2025eagle3}, and other drafter techniques~\cite{wen2024ctcdraft,cheng2024recurrentdraft,zhao2024ouroboros}, are compatible with \TheName as they all improve ETR over simple approach (e.g., Lookahead Decoding).

\subsection{Dynamic Speculation Length Selection}
\label{sec:dyn_sl}

Recent work has also explored adaptive speculation length. Dynamic Depth Decoding (DDD)~\cite{brown2024ddd} optimizes EAGLE-2 by adjusting draft tree depth based on model confidence. SVIP~\cite{zhang2024svip} introduces a self-verification policy that estimates draft length from a theoretical acceptance rate bound, while DISCO~\cite{mamou2024disco} uses a classifier to adapt speculation length per iteration. PEARL~\cite{liu2024pearl} adds pre-verify and post-verify stages for parallel drafting and verification, adaptively adjusting draft lengths to reduce idle time. However, unlike \TheName, none of these techniques focus on the data movement cost due to speculation. They require access to output probability distributions and are incompatible with approaches like n-gram speculation. Also, they rely on aggressive drafting, assuming very low over-speculation penalties (1–2\% per unit increase in K), and must draft/verify at least one token to estimate benefits. 
Moreover, such schemes introduce CPU to GPU communication between drafter iterations on the GPU, to apply policy heuristics.
Consequently, stopping criteria are used infrequently--e.g., DDD defers until the $5^{th}$ drafter iteration—making these methods too costly for MoEs.

\subsection{SLO-oriented Speculative Decoding}

Since for dense models, speculative decoding increases computation, not data movement, prior work in adaptive speculative decoding methods focus on reducing the excessive compute operations that hurt service level objectives (SLOs) in speculative decoding systems.
AdaServe~\cite{li2025adaserve} customizes SLOs through fine-grained speculative decoding, leveraging draft model logits 
to select K-value to meet individual SLO constraints while optimizing throughput. Staged Speculative Decoding~\cite{spector2023ssd} restructures speculative batches into a tree format and introduces multiple speculation stages.
SmartSpec~\cite{liu2024smartspec} profiles model and hardware pairs to build analytical model of execution and tune speculation-length to maximize "goodput" (which depends on ETR) to speculate more aggressively under low system load. Similarly, SpecServe~\cite{huang2025specserve} adaptively controls the number of tokens speculated based on a theoretical efficiency model to ensure that the system operates within the desired SLOs.

These SLO-oriented speculation techniques have two key problems: (i) they are designed for non-latency critical scenario of batch sizes that make decoding closer to compute intensive "knee" of the GPU, and (ii) they employ analytical modeling to predict model execution time, as they cater to dense models. 
Single-batch MoE serving is highly memory bound, rendering OI-centric heuristics uneffective. Moreover, analytically modeling MoE execution time would not work, as the verification time varies depending from request-to-request and even across iterations.

\subsection{MoE Decoding Latency Reduction}

To reduce latency, prior work has explored pruning and quantization for MoE models. 
Pruning removes under-utilized experts to reduce model size and data movement. For instance, MoE-Pruner~\cite{xie2024moe-pruner} employs one-shot pruning that eliminates less critical experts and utilizes knowledge distillation to recover quality loss.
Quantization reduces the precision of model weights and activations, leading to smaller model sizes and faster computation. MiLo~\cite{huang2025milo} introduces a mixture of low-rank compensators to augment highly quantized MoEs, effectively recovering accuracy loss from extreme quantization. Similarly, Atom~\cite{zhao2024atom} utilizes 4-bit weight-activation quantization.
Both pruning and quantization are orthogonal to \TheName, as the utility analysis remains applicable. Also, \TheName works with quantized models, as shown in our evaluations.

\subsection{Model Offloading and Expert Caching}


MoE models also stresses memory capacity requirement in modern GPUs, so prior work has proposed expert offloading and caching to address the issue. Cache-Conditional Experts~\cite{skliar2024cache-conditional} enhances on-device inference by introducing a cache-aware routing strategy that prioritizes the reuse of recently accessed experts. 
Similarly, MoE-Lightning~\cite{cao2025moe-lighning} proposes a CPU-GPU-I/O pipelining schedule, CGOPipe, which optimizes resource utilization. 
SiDA-MoE~\cite{du2024sida} exploits the sparsity in expert activation by dynamically offloading inactive experts to the system's main memory.
These proposals are orthogonal to \TheName as one would like to disable speculation if the costs outweigh the benefits, even if all activated experts are present in the GPU-memory cache of experts.

\section{Conclusion}

MoEs reduce computation and memory costs by activating only a subset of parameters per token. However, speculative decoding, which is effective in dense models, performs poorly on MoEs. Verifying multiple speculative tokens activates more experts, increasing parameter movement and raising verification latency by 2-3x, which often negates the effective token rate (ETR) improvements. In our evaluation across 35 model–task pairs (5 architectures, 7 workloads), speculation degrades MoE performance in 18 cases, with slowdowns of up to 1.5x.
We find that speculation utility, the ratio of benefit to cost, predicts performance reliably. With this insight, we propose \TheName, a utility-aware speculation framework that: (1) disables speculation when utility falls below 1, (2) backs off adaptively to avoid slowdown, and (3) employs hill-climbing to maximize utility. \TheName limits worst-case slowdown to 5\% and achieves 7–14\% average speedup over static-$K$ schemes, enabling practical and efficient speculation for MoEs.
\mytodo{double hyphens}
\bibliographystyle{ACM-Reference-Format}
\bibliography{references}

\end{document}